\documentclass[aps,pra,onecolumn,notitlepage,showpacs,showkeys,nofootinbib]{revtex4-1}

\usepackage[T1]{fontenc}
\usepackage[utf8]{inputenc}
\usepackage[english]{babel}
\usepackage{amssymb,amsfonts,amsmath,amsthm}
\usepackage{braket}
\usepackage{graphicx}
\usepackage{hyperref}
\usepackage{enumerate}
\usepackage{xspace}
\usepackage[normalem]{ulem}

\def\Tr{\operatorname{Tr}}
\DeclareMathOperator*{\argmin}{\operatorname{argmin}}
\def\rank{\operatorname{rank}}

\def\opt{\mathrm{opt}}
\def\w{\omega}
\def\A{\mathcal{A}}

\newcommand{\Hi}[1]{\mathcal{H}_{#1}}
\newcommand{\Fo}[1]{\mathcal{F}_{\! #1}} 
\def\S{\mathcal{S}}
\def\X{\mathcal{X}}
\def\Nset{\mathbb{N}}
\def\Cset{\mathbb{C}}
\def\sh{\mathrm{Sh}}

\def\ie{{\it i.e.,}\xspace}
\def\eg{{\it e.g.,}\xspace}
\def\etal{{\it et al.}\xspace}

\newcommand{\ketbra}[2]{\ket{#1}\bra{#2}}
\newcommand{\s}[1]{\ket{s_{#1}}}

\newtheorem{theorem}{Theorem}
\theoremstyle{definition}
\newtheorem{definition}{Definition}

\begin{document}
\title{Lossless quantum data  compression with exponential penalization: an
  operational interpretation of the quantum R\'enyi entropy}

\author{G. Bellomo\textsuperscript{1}, G.M. Bosyk\textsuperscript{2}, F. Holik\textsuperscript{2} and S. Zozor\textsuperscript{3}}
\affiliation{\textsuperscript{1}CONICET-Universidad de Buenos Aires, Instituto de Investigación en Ciencias de la Computación (ICC), Buenos Aires, Argentina}
\affiliation{\textsuperscript{2}Instituto  de Física  La Plata,  UNLP, CONICET,  Facultad  de Ciencias
  Exactas, Casilla de Correo 67, 1900 La Plata, Argentina}
\affiliation{\textsuperscript{2}CNRS, Laboratoire   Grenoblois  d'Image,   Parole,  Signal   et  Automatique
  (GIPSA-Lab), 11  rue  des Mathématiques,  38402  Saint Martin  d'Hères,
  France}


\begin{abstract}
  Based on the problem of quantum data compression in a lossless way, we present
  here  an   operational  interpretation  for   the  family  of   quantum  Rényi
  entropies. In order  to do this, we appeal to a  very general quantum encoding
  scheme   that   satisfies   a    quantum   version   of   the   Kraft-McMillan
  inequality. Then, in the standard situation, where one is intended to minimize
  the  usual average  length  of the  quantum  codewords, we  recover the  known
  results, namely that the von Neumann  entropy of the source bounds the average
  length  of the  optimal  codes. Otherwise,  we  show that  by invoking  an
    exponential    average   length,  related   to   an
    exponential  penalization  over  large  codewords,  the
  quantum R\'enyi entropies arise as the natural quantities relating the optimal
  encoding schemes  with the  source description, playing  an analogous  role to
  that of von Neumann entropy.
\end{abstract}



\maketitle


\flushbottom
\maketitle
%
%
\thispagestyle{empty}


\section*{Introduction}

One of  the main  concerns in  classical and quantum  information theory  is the
problem of encoding information by using fewest resources as possible. This task
is known  as \textit{data  compression} and it  can be  carried out either  in a
\textit{lossy}  or a \textit{lossless}  way, depending  on whether  the original
data can be recovered with or without errors, respectively.

Here, we are interested in  lossless \textit{quantum} data compression. In order
to state our proposal, let us first  recall how this task works in the classical
domain. The mathematical foundations of  classical data compression can be found
in  the seminal paper  of Shannon~\cite{Shannon1948}  (see e.g.~\cite{CoverBook}
for an introduction to the topic),  although we can summarize it as follows. Let
$S = \{p_i, s_i\}$ be a classical  source where each symbol $s_i$  has associated a
probability of occurrence $p_i$. The idea is to assign to each symbol a codeword
$c(s_i)$ of  some alphabet $A =  \{ 0 , \ldots  , k-1\}$ in an  adequate way. In
particular,  a  $k$-ary classical  code  $c$  of  $S$ is  said  \textit{uniquely
  decodable}  if this  assignment of  codewords  is injective  for any  possible
concatenation.   A celebrated  result states  that any  uniquely  decodable code
necessarily  satisfies  the \textit{Kraft-McMillan  inequality}~\cite{Kraft1949,
  McMillan1956}: $\sum_i k^{-\ell_i} \leq 1$ where $\ell_i$ is the length of the
codeword  $c(s_i)$ (measured  in bits  if $k=2$).   Conversely, given  a  set of
codewords  lengths $\{\ell_i\}$,  there exists  a uniquely  decodable  code with
these lengths.  Thus,  lossless data compression consists in  finding a uniquely
decodable code  taking into account  the statistical description of  the source.
Formally, this is  carried out by minimizing the average  codeword length $L
=\sum_i p_i \ell_i$  subject to the Kraft-McMillan inequality.   In the end, one
obtains a \textit{variable-length} code  where shorter codewords are assigned to
symbols  with a  high probability  of occurrence,  whereas larger  codewords are
assigned  to  symbols  with  low  probability  (see~\cite{CoverBook},  chap.~5).
Moreover, one  has that  (in the limit  of the  large number of  independent and
identically-distributed  sources) the  average  length of  the  optimal code  is
arbitrarily  close  to the  \textit{Shannon  entropy}~\cite{Shannon1948} of  the
source, $H(p) = - \sum_i p_i \log_k p_i$.

As noticed by  Campbell, the previous solution has the  disadvantage that it can
happen that the  codeword length turns out  to be very large for  symbols with a sufficiently
low probability  of occurrence~\cite{Campbell1965}.  Indeed, the use  of average
codewords length as a criterion  of performance has the implicit assumption that
the  cost  varies  linearly  with  the  codeword length,  which  is  not  always
desirable.  For instance, it  could be the case that adding a  letter to a large
codeword may  have a larger impact than  adding a letter to  a shorter codeword,
for instance  in terms of  memory needed to  store a codeword. This  problem has
given  place  to the  proposal  of several  other measures  of  codeword lengths  (see
e.g.~\cite{Yamano2001,  Baer2006,  Bercher2009,  Chapeau2011}),  for  which  the
average  length  is a  limiting  case.   In  particular, a  generalized  average
$t$-length, also  called exponential average,  is defined as~\cite{Campbell1965}
$L_t = \frac{1}{t} \sum_i \log_k \sum_i p_i k^{t \ell_i}$, where $t \geq 0$ is a
parameter related to  the cost. Notice that in the  limiting case $t \rightarrow
0$  one  recovers  $L_t  \rightarrow  L  $ and,  as  $t$  increases,  a  greater
\textit{penalization}  over the  large  codewords holds.   Indeed, Campbell  has
obtained a  source coding theorem taking  into account such  a penalization. His
theorem is similar to  the standard one, but the encoding is  made in such a way
that the  generalized codeword length turns  out to be arbitrarily  close to the
\textit{R\'enyi   entropy}~\cite{Renyi1961}  of   the  source,   $H_\alpha(p)  =
\frac{1}{1-\alpha} \log_k \sum_i p_i^\alpha$ with $\alpha = \frac{1}{t+1}$. This
remarkable result provides an  operational interpretation of the R\'enyi entropy
as the natural  information measure for the problem  of optimal data compression
with  penalization  over large  codewords  (see  also~\cite{Campbell1966} for  a
discussion of an axiomatic derivation of entropy related to the coding problem).

As  we  have seen,  variable-length  codes arise  naturally  in  the problem  of
lossless classical  data compression. In  the quantum information  theory realm,
the  formulation  of  this   problem  presents  intrinsic  difficulties.   These
difficulties are mainly  related to the fact that a  quantum source can possibly
send mutually non-orthogonal  states. Thus, one has to  deal with superpositions
of  quantum  codewords.  Even  worse,  these  superpositions  may correspond  to
codewords of  different lengths. Schumacher  and Westmoreland were the  first in
establishing a general approach to the problem of quantum variable-length coding
\cite{Schumacher2001}. Furthermore, they have provided the first quantum version
of the Kraft-McMillan inequality and have  found that the von Neumann entropy of
the  source $\rho$,  \ $S(\rho)  = -\Tr  \left( \rho\log_2\rho  \right)$ (binary
logarithm for coding in qubits), plays  an analogous role to that of the Shannon
entropy  in the  classical source  coding theorem.   Several other  authors have
contributed    to    this   subject    proposing    alternative   or    extended
schemes~\cite{Schumacher2001,     Bostroem2002,     Koashi2002,    Ahlswede2003,
  Ahlswede2004, Muller2008, Muller2009,  Hayashi2010, HayashiBook}.  In general,
these approaches face  the same disadvantage as in  the classical case: namely
they do  not consider  the fact  that large codewords,  even appearing  with low
probabilities,  may have  large  impact in  terms  of resources  needed for  the
encoding.  This  drawback is even more  relevant nowadays, due to  the fact that
the practical implementation of quantum information protocols pose the challenge
of manipulating coherent  superpositions of qubits.  While the  use of chains of
qubits   of  arbitrary   length  may   arise  naturally   in   some  theoretical
considerations, it can be very expensive and difficult to implement large chains
in  the  lab, specially  at  the  early stages  of  the  development of  quantum
information technology devices.  Thus, our  goal is to provide a quantum version
of  Campbell's strategy for  the problem  of coding  with penalization  of large
codewords.  As a consequence, we show that in this framework the quantum R\'enyi
entropies emerge as the natural quantities relating the optimal encoding schemes
with   the  source   description.   Accordingly,   we  provide   an  operational
interpretation for those entropies.


\section*{Results}


\subsection*{Uniquely   decodable  quantum   code  and   quantum  Kraft-McMillan
  inequality}

In this  section, we  summarize some definitions  and results of  the literature
related  to   our  proposal.    We  begin  by   pointing  out  the   problem  of
\textit{lossless quantum compression}.

Lossless quantum compression  consists in compressing a quantum  source given by
an ensemble of  quantum states, by using a variable-length  quantum code so that
the  original  states  can  be  exactly  recovered,  \ie  without  error.   More
precisely, the  situation to deal  with is the  following. Let us assume  that a
quantum source  produces an ensemble of  quantum states $\S  = \left\{p_n, \s{n}
\right\}_{n=1}^N$, where $p_n \geq 0$, $\sum_{n=1}^N  p_n = 1$ and $\s{n} \, \in
\, \Hi{\S} \, \equiv \, \Cset^d$.  The first task is to encode in an unambiguous
or uniquely  decodable way not  only every single  quantum state $\s{n}$  of the
source, but also any string of quantum states of the source.  In this sense, let
us first  introduce a  very general definition  of a  \textit{uniquely decodable
  quantum source code}.
\begin{definition}
\label{def:QuantCode}
A uniquely decodable quantum source code of $\S$ over a quantum $k$-ary alphabet
$\A =  \left\{ \ket{0} ,  \ldots , \ket{k-1}  \right\} \, \subset \,  \Hi{\A} \,
\equiv \, \Cset^k$, with $k \in  \Nset^* \backslash \{1\}$, is a linear isometry
map   \  $U:   \Fo{\S}  \rightarrow   \Fo{\A}$   \  where   \  $\Fo{\X}   \equiv
\bigoplus_{\ell=0}^{\infty} \Hi{\!   \X}^{\: \otimes \ell}$  \ is a  Fock space,
where $\X = \S$ or $\A$.
\end{definition}
In this  way, the fact that $U$  is an isometry guarantees  an injective mapping
which assigns for each string of the form \ $\otimes_{m=1}^M \s{i_m}$, with $i_m
\in  \{1  ,  \ldots ,  N\}$  and  $M  \in  \Nset^*$,  a quantum  codeword  \  $U
\otimes_{m=1}^M \s{i_m} \in \Fo{\A}$.
Let us  see how  our definition works  for single  code words. A  single quantum
codeword  over $\A$  is a  quantum pure  state that  belongs to  the  Fock space
$\Fo{\A}$ (we  are taking here strings  with a single component).   Thus, we can
write \ $U \ket{s_n} = \sum_j a_{j,n} \ket{a_{j,n}}$, where $a_{j,n} \in \Cset$,
\ $\sum_j  |a_{j,n}|^2 = 1$ \  and \ $\ket{a_{j,n}}\in  \Hi{\A}^{\: \otimes l_j}
\subset \Fo{\A}$.  Notice  that the number of non-vanishing  coefficients in the
set  $\{a_{j,n}  \}_{i=1}^{\infty}$ could  be  infinite  in  principle.  In  the
following we  will restrict to the  finite case (\ie $a_{j,n}=0$  for almost all
$j$).

Up to now, we have given  a very formal definition of uniquely decodable quantum
source   code.   In   order  to   show   an  encoding   scheme  that   satisfies
definition~\ref{def:QuantCode},   we   mainly    follow   the   proposal   given
in~\cite{Hayashi2010, HayashiBook}.   First, let us precise the  definition of a
$k$-ary classical uniquely decodable code for the symbols source $S=\{1 , \ldots
, d\}$ over  an alphabet $A =  \{ 0 , \ldots ,  k-1\}$.  Let $F_A$ be  the set \
$F_A = \bigcup_{\ell=0}^{\infty} A^{\ell}$.  Then, \ $c: S \rightarrow F_A$ \ is
a  classical uniquely  decodable code  if and  only if  for any  $M \ge  1$, any
concatenation $c^M\left(i_1,\ldots,i_M  \right) =  c(i_1) \cdots c(i_M)$  of $M$
codewords is  an injective function  (see e.g.~\cite{CoverBook}).  We  denote by
$\ell_i$ the length of the $i$-th codeword, \ie the number of ``letters'' of $A$
appearing in the codeword $c(i)$.
Hereafter, we consider  the isometries $U : \Hi{\S}  \rightarrow \Fo{\A}$ of the
form~\cite{Hayashi2010, HayashiBook}
\begin{equation}\label{eq:HayCode}
U = \sum_{i=1}^d \ketbra{c(i)}{e_i},
\end{equation}
where  $\{\ket{e_i}\}_{i=1}^d$ is  a  basis  of $\Hi{\S}$  and  $c$ a  classical
uniquely decodable code of $S$.  Clearly, by construction, one has \ $\ket{c(i)}
\in \Hi{\A}^{\: \otimes \ell_i} \,  \subset \, \Fo{\A}$ \ and \ $\{\ket{c(i)}\}$
\  forms an orthonormal  set, so  that \  $U^\dag U  = I$  (but notice  that, in
general, $U U^\dag$  can be different from the identity  operator). We refer any
isometry  of  the  form~\eqref{eq:HayCode}  as \emph{lossless  quantum  encoding
  scheme}.
Note that  contrary to a classical  code, $\ket{c(i)}$ here does  not encode any
quantum  state of  the source  $\{\ket{s_n}\}$ but  the base  state $\ket{e_i}$,
except when $\ket{s_n} = \ket{e_i}$  for some $n,i$. As introduced, the codeword
associated  to a  superposition of  source states  is the  superposition  of the
codewords.  Moreover,  $U \ket{s_n}$ does not  necessarily belong to  a space of
the form $\Hi{\A}^{\: \otimes \ell}$ for some $\ell$.  Notice now that a quantum
coding  scheme \  $U$ can  be extended  to a  map $U^M:  \Hi{\S}^{\:  \otimes M}
\rightarrow \Fo{\A}$ on sentences a follows:
\begin{equation}\label{eq:HayCodeExtended}
U^M =\sum_{i_1 = 1}^d \cdots \sum_{i_M = 1}^d
\ketbra{c(i_1) \cdots c(i_M)}{e_{i_1} \cdots e_{i_M}}.
\end{equation}
The above map is well defined for all  $M \in \Nset^*$.  A map such as $U^M$ can
be naturally  considered as an  operator acting in  the Fock space  $\Fo{\S}$ by
viewing  $\ket{e_{i_1}  \cdots  e_{i_M}}  \in  \Hi{\S}^{\:  \otimes  M}  \subset
\Fo{\S}$ as follows.  Consider a state $\ket{\phi} \in \Hi{\S}^{\: \otimes M'}$.
Then,  we  write \  $U^M  \ket{\phi} =  \delta_{M,M'}  \sum_{i_1  = 1}^d  \cdots
\sum_{i_M  = 1}^d  \braket{e_{i_1}  \cdots e_{i_M}  |  \phi} \ket{c(i_1)  \cdots
  c(i_M)}$.  Now, with this observation we can define an operator \ $U^{\infty}:
\Fo{\S} \rightarrow \Fo{\A}$ \ as
\begin{equation}
U^{\infty} = \sum_{M = 1}^{\infty} U^M.
\end{equation}
The  physical   interpretation  of  $U^{\infty}$  is  that   for  each  sentence
$\otimes_{m=1}^M \ket{s_{i_m}}$  of the source,  we will obtain the  right coded
sentence for  each $M \in  \Nset^*$.  It is  important to remark that  all these
coding schemes are lossless in the sense of definition~\ref{def:QuantCode}.


As it is well known in classical data compression, the Kraft-McMillan inequality
gives  a necessary  and sufficient  condition for  the existence  of  a uniquely
decodable  code  (see \eg\cite{CoverBook}).   This  result  has been  originally
extended   to  the  quantum   domain  in~\cite{Schumacher2001},   introducing  a
particular formalism.   We proceed here  to obtain a \textit{quantum  version of
  the Kraft-McMillan inequality}, compatible with the previous construction.

Let us  first introduce  the \textit{length observable},  which allows to  get a
further notion of codeword length.
\begin{definition}
\label{def:lobs}
The length observable $\Lambda$ acting on $\Fo{\A}$ is defined as
\begin{equation}\label{eq:lobs}
\Lambda \equiv \sum_{\ell=0}^{\infty} \ell \, \Pi_\ell,
\end{equation}
where $\Pi_\ell$ denotes the orthogonal projector onto the subspace $\Hi{\A}^{\:
  \otimes \ell} \subset \Fo{\A}$.
\end{definition}

Now, the quantum Kraft-McMillan inequality reads as follows.
\begin{theorem}
  \label{th:QKMI}  For  any  losless   quantum  encoding  scheme  $U$  given  by
  Eq.~\eqref{eq:HayCode}, the following inequality must be satisfied:
\begin{equation}
\label{eq:QKMI}
\Tr \left(  U^\dag k^{- \Lambda}  U \right) \leq 1.
\end{equation}
\end{theorem}
\noindent The proof of this theorem, which mainly relies in its classical counterpart, is given in the section Methods, along with the proofs of the subsequent theorems.


\subsection*{Source coding and von Neumann entropy bounds}


As in the  classical case, we are interested in quantum  codes that minimize the
amount  of resources involved.   However, in  the quantum  case arises  an extra
difficulty to quantify the number of resources since there is no a unique way of
defining the notion of length of a quantum codeword. For a given encoding scheme
$U$,  the  standard  definition  of  \textit{quantum  codeword  length}  is  the
following.
\begin{definition}
\label{def:lengthcodeword}
The quantum codeword length of $\ket{\w} \equiv U \ket{s}$ for some $\ket{s} \in
\Hi{\S}$ is given by the expectation value
\begin{equation}
\label{eq:lengthcodeword}
\ell(\ket{\w}) \, \equiv \, \, \braket{\w | \Lambda | \w}\, = \sum_{i=1}^d
\left| \braket{e_i | s} \right|^2 \ell_i.
\end{equation}
\end{definition}
Thus, from  this definition, the codewords  may not have definite  length in the
sense that they are not eigenstates  of the length operator in the general case.
For that reason  a quantum code given by  the encoding scheme~\eqref{eq:HayCode}
is       sometimes      called      {\em       quantum      indeterminate-length
  code}~\cite{Schumacher2001}.

As we have noticed, one can introduce another important measure of the length of
a  quantum   codeword.   One  used   in  the  literature  is   the  \textit{base
  length}~\cite{Bostroem2002}:
\begin{definition}
\label{def:baselength}
The base length of a quantum codeword $\ket{\w} \equiv U \ket{s}$ is given by
\begin{equation}
\label{eq:baselength}
l(\ket{\w}) \, \equiv \, \max \left\{\ell \in \Nset \, | \, \braket{\w |
\Pi_\ell | \w} \neq 0 \right\} \:\:\: = \max_{\{ i \in \{1, \ldots, d\}|
\braket{e_i | s} \neq 0\}} \{\ell_i \}.
\end{equation}
\end{definition}
Notice that the base  length plays a key role as it  determines the minimum size
of the quantum register necessary to store a quantum codeword.

The base length of a quantum codeword is an integer whereas the quantum codeword
length is  not, in general.  However,  there is a relation  between both lengths
given  by  $\ell(\ket{\w})  =  \sum_{\ell =  0}^{l(\ket{\w})}  \ell  \braket{\w|
  \Pi_\ell  |  \w}  \le  l(\ket{\w})  \sum_\ell  \braket{\w|  \Pi_\ell  |  \w}$.
Immediately,  one has $\ell(\ket{\w})  \leq l(\ket{\w})$,  with equality  if and
only if $\ket{\w} = U \ket{s}$ is  an eigenstate of $\Lambda$ , \ie if $\ket{s}$
is an eigenstate of $U$.

Henceforth, we  consider that the state of  the quantum source $\S$  is given by
the density operator $\rho$, \ie  a positive semi-definite operator of trace one
acting on $\Cset^d$. We will  write the density operator using the decomposition
on  ensemble's states,  \ie $\rho  = \sum_{n=1}^{N}  p_n  \ketbra{s_n}{s_n}$, or
equivalently, considering the spectral decomposition, \ie $\rho = \sum_{i=1}^{d}
\rho_i \ketbra{\rho_i}{\rho_i}$, where  $\rho_i$ is the eigenvalue corresponding
to the eigenstate $\ket{\rho_i}$. In addition, we will denote as
\begin{equation}
\label{eq:output}
C(\rho) \equiv U \rho U^\dag = \sum_{i,i'=1}^{d} \braket{e_i |\rho |e_{i'}}
\ketbra{c(i)}{c(i')}
\end{equation}
the output of the quantum encoder~\eqref{eq:HayCode}.
Then,  according  to  definition~\ref{def:lengthcodeword},  the  \textit{average
  codeword length} of $\S$ is given by
\begin{equation}
\label{eq:avgcodeword}
\ell(C(\rho)) \, \equiv \, \, \Tr \left( C(\rho) \Lambda \right) \, =
\sum_{n=1}^N p_n \sum_{i=1}^d \left| \braket{e_i|s_n} \right|^2 \ell_i.
\end{equation}
On   the   other  hand,   according   to  definition~\ref{def:baselength},   the
\textit{base length} of $\S$ is
\begin{equation}
\label{eq:baselengthsource}
l(C(\rho)) \, \equiv \, \max \{l(U \s{n})\}_{n=1}^N = \max \left\{\max_{\{ i \in
\{1, \ldots, d\}| \braket{e_i | s_n} \neq 0\}} \{\ell_i \} \right\}_{n=1}^N.
\end{equation}

We have  now all the  ingredients to introduce \textit{optimal  quantum lossless
  codes}.

\begin{definition}
\label{def:optcod}
A quantum encoding scheme $U$ is optimal for the source $\S$ if it minimizes the
average codeword length, that is,
\begin{equation}
\label{eq:optcod}
U^\opt \equiv \argmin_{\Tr \left( U^\dag k^{-\Lambda} U \right) \leq 1} \Tr
\left(U \rho U^\dag \Lambda \right)
\end{equation}
and thus the minimal average codeword length for the source $\S$ is given by
\begin{equation}
\label{eq:minavlength}
\ell(C^\opt(\rho)) \, = \, \Tr \left( C^\opt(\rho)  \Lambda  \right),
\end{equation}
where  $C^\opt(\rho) \equiv U^\opt \rho {U^\opt}^\dag$.
\end{definition}

In the classical setting to search for the optimal code, one has to find for the
set of integers $\{\ell_i\}$ that minimizes the averaged length subjected to the
Kraft-McMillan inequality. It is  well known that \textit{Huffman code} provides
the optimal  solution \cite{Huffman1952}.  Let  us see that the  quantum optimal
code or the  quantum version of Huffman code is obtained  for an encoding scheme
$U$ with  basis given by  the eigenstates of  $\rho$ and the classical  code $c$
given  by  the  Huffman  code  for  the  symbols  $\{1  ,  \ldots  ,  d\}$  with
probabilities given by the eigenvalues of $\rho$.
\begin{theorem}
\label{th:optcod}
The optimal quantum code of the quantum source $\S$ writes
\begin{equation}
\label{eq:optimalcode}
U^\opt = \sum_{i=1}^d \ketbra{c^\opt(i)}{\rho_i},
\end{equation}
where  $\{ c^\opt(i)  \}$ is  the classical  optimal code  given by  the Huffman
code~\cite{Huffman1952} of the  symbols $\{1 , \ldots ,  d\}$ with corresponding
probabilities $\{ \rho_1 , \ldots , \rho_d \}$.
\end{theorem}

Let us  recall that there is no  an analytic formula for  the individual lengths
$\ell_i$ of the  classical Huffman code in the general case.  On the other hand,
if  one  drops the  integer  restriction  of  $\{\ell_i\}$ in  the  minimization
problem, one obtains the optimum  ``lengths'' $-\log_k \rho_i$.  To take integer
values, one  can consider  the excess  integer part of  these values,  $\ell_i =
\lceil  -\log_k \rho_i  \rceil$, and  construct a  corresponding code  using the
Kraft tree (see\cite{CoverBook} for more  details).  This is a well known method
called the \textit{Shannon coding} for which  the average length is close to the
optimal one (which  is given by the Huffman code). Accordingly,  we can say that
the quantum version  of the Shannon code  is given by an encoding  scheme of the
form \eqref{eq:optimalcode},  where the classical code  $c$ is now  given by the
Shannon code.  Nevertheless, without  explicitly expressing the optimal code, it
is possible  to upper  and lower  bound the optimal  average codeword  length in
terms  of the  von Neumann  entropy of  the source,  as previously  proved  in a
different formalism in~\cite{Schumacher2001}.
\begin{theorem}
\label{th:bounds}
The average length of the optimal code is lower and upper bounded as follows
\begin{equation}
\label{eq:bounds}
S(\rho) \leq \ell(C^\opt(\rho)) < S(\rho) +1,
\end{equation}
where  $S(\rho) = -  \Tr \left(  \rho \log_k  \rho \right)$  is the  von Neumann
entropy of a density operator $\rho$, and $\log_k$ is the logarithm of base $k$.
\end{theorem}

According  to theorem~\ref{th:bounds},  the  entropy of  the  source bounds  the
compression capacity.  Moreover, one can attain  the lower bound for the case of
$K$ independent  and identical  preparations of the  source for large  $K$.  Let
$\rho^{\otimes  K}$  be  the  corresponding  density  operator,  and  denote  by
$\frac{1}{K} \, \ell(C^\opt(\rho^{\otimes K}))$  the optimal average length code
per    source,    where   $C(\rho^{\otimes    K})$    is    defined   via    the
concatenation~\eqref{eq:HayCodeExtended}.  Then,  from $S(\rho^{\otimes K})  = K
S(\rho)$ and theorem~\ref{th:bounds}  one has $S(\rho) \, \le  \, \frac{1}{K} \,
\ell(C^\opt(\rho^{\otimes K})) \, < \, S(\rho) + \frac{1}{K}$, so that
\begin{equation}
\lim_{K \to \infty} \, \frac{1}{K} \, \ell(C^\opt(\rho^{\otimes K})) \, = \,
S(\rho).
\end{equation}

We end this section discussing what  happens to the average codeword length when
the encoding scheme is designed  for a ``wrong'' density operator $\tau$ instead
of the correct one $\rho$. This could be useful for the case where $\tau$ is the
best estimation of  the state of the source for instance.   In such a situation,
the average code  length of the quantum Shannon code  corresponding to $\tau$ is
again bounded, as follows (see, \eg~Refs.\cite{Hayashi2010,HayashiBook}).
\begin{theorem}
\label{th:wrongcode}
Let $\tau$  be a density  operator whose diagonal  form is $\tau  = \sum_{i=1}^d
\tau_i \ketbra{\tau_i}{\tau_i}$. Let us consider the quantum Shannon code $U^\sh
=  \sum_{i=1}^d \ketbra{c(i)}{\tau_i}$  designed  for $\tau$,  where $c(i)$  are
classical codewords of the Shannon code,  with lengths $\ell_i = \lceil - \log_k
\tau_i \rceil$.   The average length  of such a  quantum encoding is  bounded as
follows
\begin{equation}
\label{eq:wrongcode}
S(\rho) + S(\rho \| \tau) \leq \ell(C^\sh(\rho)) < S(\rho) + S(\rho \| \tau) + 1,
\end{equation}
where $C^\sh(\rho) \equiv U^\sh \rho {U^\sh}^ \dag$.
\end{theorem}
Notice that  this gives  an operational interpretation  to the  quantum relative
entropy as  follows: $S(\rho \| \tau)$  measures the deviation  from the average
codeword length of  the quantum Shannon code, when the code  is designed using a
density operator  which differs from  density operator associated to  the source
(see also  \cite{SchumacherBook, Kaltchenko2008} for a  further understanding of
the role of quantum relative entropy in the context of data compression).


\subsection*{Source coding and quantum R\'enyi entropy bounds}

Let us first  note that the definition~\ref{def:optcod} of  optimal code and the
results    given     above    are    closely    linked     to    the    standard
definition~\ref{def:lengthcodeword}    of    the    length    of    a    quantum
codeword. However,  there could  be problems for  which the relevant  measure of
length  is not  the usual  one.  In  this sense,  M\"uller \etal  have  used the
average of the base lengths of the source in order to define a different optimal
code~\cite{Muller2009} and have obtained a complementary result to the one given
by theorem \ref{th:bounds}.  In this  section we follow an alternative strategy,
which  is  based  on  an   extension  of  Campbell's  proposal  to  the  quantum
case~\cite{Campbell1965}.     Let    us   first    introduce    a   notion    of
\textit{exponential} quantum codeword length.  The standard quantum codeword and
base lengths turn out to be particular cases of our definition.
\begin{definition}
\label{def:t-lengthcodeword}
The $t$-exponential length of a quantum codeword $\ket{\w} \equiv U \ket{s}$ for
some $\ket{s} \in \Hi{\S}$ is given by the expectation value
\begin{equation}
\label{eq:t-lengthcodeword}
\ell_t(\ket{\w}) \, \equiv \, \, \frac{1}{t} \log_k \braket{\w|k^{t \Lambda} |
\w} \, = \, \frac{1}{t} \log_k \left( \sum_{i=1}^d \left| \braket{e_i|s}
\right|^2 k^{t \ell_i} \right),
\end{equation}
where  $t  \geq  0$ is  a  parameter  related  to  the  cost assigned  to  large
codewords. In the limiting cases, one has
\begin{equation}
\label{eq:lengthlimits}
\ell_0(\ket{\w}) \equiv \lim_{t \rightarrow 0} \ell_t(\ket{\w}) = \ell(\ket{\w})
\qquad \mbox{and} \qquad \ell_{\infty}(\ket{\w}) \equiv \lim_{t \rightarrow
\infty} \ell_t(\ket{\w}) = l(\ket{\w}).
\end{equation}
\end{definition}
Notice that $t \mapsto \ell_t(\ket{\w})$ is a continuous nondecreasing function,
\ie  $\ell_t(\ket{\w}) \leq  \ell_{t'}(\ket{\w})$  for $t  \leq  t'$.  Thus,  by
changing the parameter $t$, one  can move continuously and increasingly from the
standard  quantum  codeword length  to  the base  length.  In  other words,  the
$t$-exponential  codeword  length  will  allow  to  make  a  compromise  between
minimizing  the  average length  and  the base  length.  Finally,  note that  if
$\ket{\w}  \in \Cset^\ell$, \ie  the quantum  codeword is  an eigenstate  of the
length  observable,  then  $\ell_t(\ket{\w})  =  \ell$, which  is  a  reasonable
property for a quantum codeword length measure.

According to  definition~\ref{def:t-lengthcodeword}, the $t$-exponential average
codeword length of the quantum source $\S$ is given by
\begin{equation}
\label{eq:t-avgcodeword}
\ell_t(C(\rho)) \, \equiv \,\frac{1}{t} \log_k \Tr \left( C(\rho) \,
k^{t\Lambda} \right) \, = \, \frac{1}{t} \log_k \left( \sum_{n=1}^N p_n
\sum_{i=1}^d \left| \braket{e_i|s_n} \right|^2 k^{t\ell_i} \right).
\end{equation}

We  introduce  now the  notion  of optimal  quantum  code  corresponding to  our
previously  defined $t$-exponential  codeword  length. A  natural  choice is  as
follows:
\begin{definition}
\label{def:t-optcod}
A quantum encoding scheme $U$ is  $t$-exponential optimal for the source $\S$ if
it minimizes the $t$-exponential average codeword length, that is,
\begin{equation}
\label{eq:t-optcod}
U^\opt_t \equiv \argmin_{\Tr \left( U^\dag k^{-\Lambda} U \right) \leq 1}
\frac{1}{t} \log_k \Tr \left( U \rho U^\dag k^{t \Lambda} \right)
\end{equation}
and thus the minimal $t$-exponential average codeword length for the source $\S$
is given by
\begin{equation}
\label{eq:t-minavlength}
\ell_t(C^\opt_t(\rho)) \, = \, \frac{1}{t} \log_k \Tr \left( C^\opt_t(\rho) k^{t
\Lambda} \right),
\end{equation}
where $C^\opt_t(\rho) \equiv U^\opt_t \rho {U^\opt_t}^\dag$.
\end{definition}

In the classical setting to search  for the $t$-exponential optimal code, as for
the standard  context, one has to  look for the  set of integers $\{  \ell_i \}$
that   minimizes  the   $t$-exponential   averaged  length   subjected  to   the
Kraft-McMillan   inequality.     This   problem   has    been   already   solved
in~\cite{Parker1980, Humblet1981,  Baer2006}.  In the quantum  context, we prove
here that  the optimal code  is again obtained  for an encoding scheme  $U$ with
basis  given by  the eigenstates  of  $\rho$ and  the classical  $t$-exponential
optimal code $c_t$ for the symbols $\{1 , \ldots , d\}$ with probabilities given
by the eigenvalues of $\rho$.
\begin{theorem}
\label{th:toptcod}
The quantum code  that minimizes the $t$-exponential average  codeword length of
the quantum source $\S$ writes
\begin{equation}
\label{eq:toptimalcode}
U_t^\opt = \sum_{i=1}^d \ketbra{c_t^\opt(i)}{\rho_i},
\end{equation}
where $\{ c_t^\opt(i)  \}$ is the classical code  minimizing the $t$-exponential
average  code length  of the  symbols $\{1  , \ldots  , d\}$  with corresponding
probabilities $\{ \rho_1 , \ldots , \rho_d \}$.
\end{theorem}

As for  the standard case,  there is no  an analytic formula for  the individual
optimal integer lengths $\ell_i$  leading to the minimum $t$-exponential average
length of the  classical code. But, again, if one  drops the integer restriction
of  $\{ \ell_i  \}$ in  the minimization  problem, one  obtains now  the optimum
``lengths''  $-\log_k  {\rho_t}_i$  where  the  ${\rho_t}_i$  are  the  ``escort
probabilities'', eigenvalues of the ``escort'' density operator
\begin{equation}
\label{eq:rhot}
\rho_t \equiv \frac{\rho^{\frac{1}{1+t}}}{\Tr\rho^{\frac{1}{1+t}}},
\end{equation}
acting on $\Hi{\S}$.  To take integer  values, one can again consider the excess
integer part of these values, $\ell_i  = \lceil - \log_k {\rho_t}_i \rceil$, and
construct a  corresponding code using the  Kraft tree, that is  the Shannon code
corresponding  to  the  escort   probabilities  $\{  {\rho_t}_i  \}$.   However,
independently  of  the  explicit  expression  of the  generalized  optimal  code
\eqref{eq:t-optcod},  it  is possible  to  upper  and  lower bound  the  optimal
$t$-exponential  average  quantum  codeword  length~\eqref{eq:t-minavlength}  in
terms  of the  quantum R\'enyi  entropy of  the  source.
\begin{theorem}
\label{th:t-bounds}
The $t$-exponential average length of  the $t$-exponential optimal code is lower
and upper bounded as follows
\begin{equation}
\label{eq:t-bounds}
S_{\frac{1}{1+t}}(\rho) \leq \ell_t(C^\opt_t(\rho)) < S_{\frac{1}{1+t}}(\rho) +1,
\end{equation}
where  $S_\alpha(\rho) = \frac{1}{1-\alpha}  \log_k \Tr  \rho^\alpha, \: \alpha
  \ge 0$, is the quantum R\'enyi  entropy of the density operator of the source
$\rho$.
\end{theorem}

We recall that our aim is to provide a scheme to address the problem of how to codify codewords of a quantum source allowing chains of variable length, but considering a penalization for large codewords.
This aim can be achieved by appealing to definitions~\ref{def:t-lengthcodeword} and~\ref{def:t-optcod} and theorems~\ref{th:toptcod} and~\ref{th:t-bounds}.
In particular, we can interpret theorem~\ref{th:t-bounds} as the quantum version of Campbell’s source coding theorem~\cite{Campbell1965}. Hence, the quantum Rényi entropy plays a role similar to that of von Neumann’s in the standard quantum source coding theorem, when an exponential penalization is considered. Indeed, theorem~\ref{th:bounds} results as a particular case of our theorem~\ref{th:t-bounds} (with $t=0$), recovering the results of Schumacher and Westmoreland~\cite{Schumacher2001}. This situation is completely analogous to that of the classical setting, with regard to the roles played by Rényi and Shannon measures for the cases with and without penalization, respectively. Consequently, this allows us to provide a natural operational interpretation for the quantum Rényi entropy in relation with the problem of lossless quantum data compression. Finally, notice that this is an alternative approach to that of Müeller et al.~\cite{Muller2009}, where they have studied an analogous problem, but minimizing the average of the individual base lengths of the source instead of considering a penalization over large codewords.

According  to  theorem~\ref{th:t-bounds}, the  quantum  R\'enyi  entropy of  the
source  bounds the  compression  capacity when  an  exponential penalization  is
considered.  As in the case with no penalization, one can attain the lower bound
for  the case  of $K$  independent and  identically prepared  sources  for large
$K$.  Thus,  consider  a  density  operator $\rho^{\otimes  K}$  and  denote  by
$\frac{1}{K}  \,  \ell_t(C^\opt_t(\rho^{\otimes  K}))$  to  the  $t$-exponential
optimal    average    length    code    per   source.     Then,    using    that
$S_\alpha(\rho^{\otimes K})  = K S_\alpha(\rho)$  and theorem~\ref{th:t-bounds},
one has  $S_{\frac{1}{1+t}}(\rho) \leq \frac{1}{K} \ell_t(C^\opt_t(\rho^{\otimes
  K})) \, < \, S_{\frac{1}{1+t}}(\rho) + \frac{1}{K} $. In this way
\begin{equation}
\lim_{K \to \infty} \, \frac{1}{K} \ell_t\left( C^\opt_t(\rho^{\otimes K})
\right) \, = \, S_{\frac{1}{1+t}}(\rho).
\end{equation}
Let us point out that the quantum Rényi entropy appears also naturally in the determination of the exponent of the average error of the quantum fixed-length source coding~\cite{Hayashi2002,HayashiBook2}, which is closely related to the Chernoff exponent appearing in classical discrimination problems. This exponent provides thus another interpretation of the quantum Rényi entropy. Our approach differs in that we study the role played by the quantum R\'enyi entropy in the problem of lossless quantum data compression with penalization.

As in  the end of  the previous  section, we now  discuss what happens  with the
$t$-exponential average codeword length when the encoding scheme is designed for
a density operator $\tau$, \ie  using the escort density operator $\tau_t \equiv
\frac{\tau^{\frac{1}{1+t}}}{\Tr   \tau^{\frac{1}{1+t}}}$.   In  that   case  the
$t$-exponential   average  codeword   length   of  the   quantum  Shannon   code
corresponding to $\tau_t$ is again bounded as follows.

\begin{theorem}
\label{th:t-wrongcode}
Let $\tau$  be a density  operator whose diagonal  form is $\tau  = \sum_{i=1}^d
\tau_i  \ketbra{\tau_i}{\tau_i}$.   Let us  consider  the  quantum Shannon  code
$U^\sh_t =  \sum_{i=1}^d \ketbra{c(i)}{\tau_i}$ designed for  the escort density
operator $\tau_t$,  where $c(i)$  are classical codewords  of the  Shannon code,
with lengths $\ell_i = \lceil  - \log_k {\tau_t}_i \rceil$.  The $t$-exponential
average length of such a quantum encoding is bounded as follows
\begin{equation}
\label{eq:t-wrongcode}
S_{\frac{1}{1+t}}(\rho) + S_{1+t}(\rho_t \| \tau_t) \leq \ell_t(C^\sh_t(\rho)) <
S_{\frac{1}{1+t}}(\rho) + S_{1+t}(\rho_t \| \tau_t) + 1,
\end{equation}
where $C^\sh_t(\rho) \equiv U^\sh_t \rho {U^\sh_t}^\dag$.
\end{theorem}
It  is   important  to  remark   that  this  theorem  provides   an  operational
interpretation for  the quantum R\'enyi divergence as  follows: $S_{1+t}( \rho_t
\| \tau_t)$  quantifies the deviation from the  $t$-exponential average codeword
length of  the quantum Shannon code, when  the code is designed  using an escort
density operator which differs from density operator associated to the source.

It would  be desirable to have  some expression that indicates  how the standard
average and the  base length of the $t$-exponential optimal  code behave when an
exponential penalization is  considered. However, this is not  possible as there
is no  an analytic formula for the  individual codeword length in  this case, in
general. An interesting alternative is analyzing how the the standard
average of the quantum Shannon code is
affected  by  an  exponential   penalization.

\begin{theorem}
\label{th:tradeoff}
Let  $U^{\sh}_t  =  \sum_{i=1}^d
\ketbra{c(i)}{\rho_i}$  be the  quantum  Shannon code  designed  for the  escort
density operator $\rho_t$,  for which the classical codewords  lengths are given
by $\ell_i = \left\lceil - \log_k {\rho_t}_i \right\rceil$.
The average length of this code is bounded as follows
\begin{equation}
\label{eq:averageUmin}
\frac{1}{1+t} S(\rho) + \frac{t}{1+t} S_\frac{1}{1+t}(\rho) \, \le \,
\ell(C^\sh_t(\rho)) \, < \, \frac{1}{1+t} S(\rho) + \frac{t}{1+t}
S_\frac{1}{1+t}(\rho) + 1.
\end{equation}
\end{theorem}

Notice that  the bounds are  basically a convex  combination of the  von Neumann
entropy (related to the minimum average length) and the R\'enyi entropy (related
to  the  minimal  $t$-exponential  average  length)  of  the  source.   Since  $
S_\frac{1}{1+t}(\rho)  \geq S(\rho)$,  the average  length $\ell(C^\sh_t(\rho))$
increases  with  respect  to   $t$;  in  particular,  $\ell(C^\sh_t(\rho))  \geq
\ell(C^\sh(\rho))$. On the contrary, for the  base length, one can see that when
$\rho_i$ is  small enough,  there exists a  parameter $t$ sufficiently  large so
that $l(C^\sh_t(\rho)) < l(C^\sh(\rho))$.  In particular, the base length can be
lessen up  to $\lceil  S_0(\rho)\rceil = \lceil  \log_k \rank \rho  \rceil$. So,
there is  a tradeoff  between $\ell(C^\sh_t(\rho))$ and  $l(C^\sh_t(\rho))$ with
respect  to $t$.   The optimal  choosing of  the cost  parameter depends  on the
particularities  of the  problem  in question  (e.g.,  the size  of the  quantum
register, etc).  Finally, notice that for the exceptional case that all $-\log_k
{\rho_t}_i$ are integers, the quantum Shannon code hence designed coincides with
the $t$-exponential optimal code  $C_t^\opt$ of theorem~\ref{th:toptcod} and the
lower bound of~\eqref{eq:t-bounds} is achieved.


\section*{Discussion}

We have addressed the problem of lossless quantum data compression.
In particular, we have considered the case in which codification of
large codewords is penalized. Our work can be regarded as a quantum
version of Campbell's work~\cite{Campbell1965}.

First, we have provided an expression for the optimal code for the
case with exponential penalization (theorem~\ref{th:toptcod}) in
terms of its classical counterpart~\cite{Parker1980, Humblet1981,
Baer2006}. We have shown that this penalization affects the optimal
code in such a way that the R\'enyi entropy of the source bounds the
$t$-exponential average codeword length (theorem~\ref{th:t-bounds}).
As a corollary, in the limit of a large number of independent and
identically prepared sources, we have found that the capacity of
compression equals the R\'enyi entropy of the source. Thus, the
quantum R\'enyi entropy acquires a natural operational
interpretation. In addition, we have found that a wrong description
of the source produces an excess term in the bound of the average
codeword length, which is related to the quantum R\'enyi divergence
(theorem~\ref{th:t-wrongcode}). Given that we recover the results by
Schumacher and Westmoreland~\cite{Schumacher2001} when penalization
is negligible, our work can be seen as an generalization of theirs.

Finally, we have discussed how the average and base lengths of the
quantum Shannon code behave in terms of the cost parameter, which is
related to the penalization (theorem~\ref{th:tradeoff}). Indeed, there is a tradeoff between
these two quantities, in the sense that it is possible to reduce the
base length, but with the side effect of increasing the average
length and viceversa.

It is worth noticing that our approach provides an alternative to
that of M\"ueller \etal~\cite{Muller2009}, where they have studied
an analogous problem, but minimizing the average of the individual
base lengths of the source. Our results are complementary to theirs.


\section*{Methods}

In this section, we give the proofs of all theorems.

Proof of theorem~\ref{th:QKMI}.

\begin{proof}
  Notice  first   that  $U^\dag   k^{-\Lambda}  U  =   \sum_{i=1}^d  k^{-\ell_i}
  \ketbra{e_i}{e_i}$ due to
  \begin{equation}
  \bra{c(i)}   \Pi_\ell    \ket{c(i')}   =
  \delta_{\ell,\ell_{i'}}  \, \delta_{i,i'},
  \end{equation}
  where the $\ell_i$  are the lengths of the  classical codewords $c(i)$.  Then,
  one directly obtains $\Tr \left( U^\dag k^{- \Lambda} U \right) = \sum_{i=1}^d
  k^{-\ell_i}$.  Given that  the code $c$ is uniquely  decodable, the proof ends
  by appealing to the classical Kraft-McMillan inequality.
\end{proof}


Proof of theorem~\ref{th:optcod}.

\begin{proof}
  Let us first notice that  the quantum Kraft-McMillan constraint is independent
  of the basis $\{\ket{e_i}\}$ of a  given lossless quantum encoding scheme $U =
  \sum_i \ketbra{c(i)}{e_i}$.  Then,  to prove the theorem one can  do it in two
  steps.  On the  one hand, let us  first fix a classical code  $c$ and minimize
  $\ell(C(\rho))  = \sum_{i,j} \ell_i  \rho_j |\braket{e_i|\rho_j}|^2$  over the
  set of basis of $\Cset^d$.  Let  us introduce the doubly stochastic matrix $D$
  with entries  $D_{i,j} \equiv  |\braket{e_i|\rho_j}|^2$, \ie $D_{i,j}  \geq 0$
  and  $\sum_j D_{i,j}  = \sum_i  D_{i,j} =  1$  for all  $i$ and  $j$.  So  the
  minimization problem  consists in minimizing  $\vec{\ell} \ D  \ \vec{\rho}^t$
  over the set  of doubly stochastic matrices, where  $\vec{\ell} = \left[\ell_1
    \ldots  \ell_d  \right]$  and   $\vec{\rho}  =  \left[\rho_1  \ldots  \rho_d
  \right]$.  From the  Birkhoff theorem~\cite{Birkhoff1946, Bhatia1997}, one can
  write $D= \sum_k \pi_k \Pi_k$ as a convex combination of permutations matrices
  $\Pi_k$.  Thus,  $\vec{\ell} \  D \ \vec{\rho}^t  = \sum_k \pi_k  \vec{\ell} \
  \Pi_k \vec{\rho}^t \geq \vec{\ell} \  \Pi_{k'} \vec{\rho}^t$ for some $k'$, so
  that $D= \Pi_{k'}$. Although one  does not know such permutation, this implies
  that   each  element   of   $\{\ket{e_i}\}$  coincides   with   only  one   of
  $\{\ket{\rho_j}\}$. On the  other hand, one can skip  the search of $\Pi_{k'}$
  since one has now  to minimize the averaged length with respect  to the set of
  lengths  $\{\ell_i\}$  subject  to  the classical  Kraft-McMillan  inequality.
  Indeed, without loss of generality, the permutation can be incorporated in the
  lengths   by  replacing   $\vec{\ell'}  \rightarrow   \vec{\ell}   \  \Pi_k'$.
  Therefore, one has $\ket{e_i} = \ket{\rho_i}$ and thus $\ell(C(\rho)) = \sum_i
  \ell_i  \rho_i$  is  the  classical  average  length  of  the  classical  code
  $c$. Finally, one  has to find the classical optimal  code $c$, whose solution
  is well known in the literature given by the Huffman code~\cite{Huffman1952}.
\end{proof}


Proof of theorem~\ref{th:bounds}.

\begin{proof}
  Let us first introduce the density operator
  \begin{equation}\label{eq:sigmabeta}
  \sigma \, \equiv \, \frac{U^\dag k^{-\Lambda} U}{\beta} \qquad \mbox{with}
  \qquad \beta \, \equiv \, \Tr \left( U^\dag k^{-\Lambda} U \right)
  \end{equation}
  acting on $\Hi{\S}$. Let  \ $C(\rho) = U \rho U^\dag$ \  with $U$ an arbitrary
  encoding scheme of the form  \eqref{eq:HayCode}.  Then, noting that $\Lambda =
  - \log_k  k^{-\Lambda}$, and  thus that  $U^\dag \Lambda U  = -  \log_k \left(
    U^\dag k^{-\Lambda} U \right)$, it is straightforward to show that
  \begin{equation}
  \label{eq:avgent}
  \ell(C(\rho)) = S(\rho) + S(\rho \| \sigma) - \log_k \beta,
  \end{equation}
  where \ $S(\rho \| \sigma) = \Tr\left[ \rho \left( \log_k \rho - \log_k \sigma
    \right) \right]$  \ is the  quantum relative entropy.  The  quantum relative
  entropy being definite positive, and from \ $\log_k \beta \leq 0$ \ due to the
  quantum Kraft-McMillan inequality, it follows that
  \begin{equation}
  \label{eq:lowerbound}
  \ell(C(\rho)) \geq  S(\rho),
  \end{equation}
  for any encoding scheme $U$, in particular for the optimum one.

  In order to proof the upper bound,  let us consider the quantum Shannon code \
  $U^\sh = \sum_{i=1}^d \ketbra{c(i)}{\rho_i}$ \ of $\rho$, where the lengths of
  the codewords  $\left\{ c(i) \right\}$ are  $\left\{ \ell_i =  \lceil - \log_k
    \rho_i  \rceil  \right\}$.  Notice  that  this  code  satisfies the  quantum
  Kraft-McMillan inequality~\eqref{eq:QKMI} by  construction. Then, from $\lceil
  - \log_k \rho_i \rceil < - \log_k \rho_i + 1$, we have
  \begin{equation}
  \label{eq:avgcbar}
  \ell(C^\sh(\rho)) = \sum^d_{i=1} \rho_i \lceil - \log_k \rho_i \rceil < S(\rho) + 1.
  \end{equation}
  The upper bound in~\eqref{eq:bounds}  immediately follows from this inequality
  and  from $\ell(C^\opt(\rho))  \leq \ell(C^\sh(\rho))$  (by definition  of the
  optimal code).
\end{proof}


Proof of theorem~\ref{th:wrongcode}.

\begin{proof}
  Let $U^\sh  = \sum_{i=1}^d \ketbra{c(i)}{\tau_i}$ be the  quantum Shannon code
  of  $\tau$.  It is  straightforward  to  show  that $\ell(C^\sh(\rho))  =
  \sum_{i=1}^d  \braket{\tau_i|\rho|\tau_i} \lceil -  \log_k \tau_i  \rceil$.  \
  The bounds  result thus  directly from  $- \log_k \tau_i  \le \lceil  - \log_k
  \tau_i \rceil < - \log_k \tau_i + 1$ and $- \sum_{i=1}^d \braket{\tau_i|\rho |
    \tau_i}  \log_k \tau_i  = -  \Tr(\rho  \log_k \tau)  = S(\rho)  + S(\rho  \|
  \tau)$.
\end{proof}


Proof of theorem~\ref{th:toptcod}.

\begin{proof}
  For a given lossless quantum encoding scheme \ $U = \sum_i \ketbra{c(i)}{e_i}$
  \ it  is straightforward  to see that  \ $\ell_t(C(\rho)) =  \frac{1}{t} \log_k
  \left(  \sum_{i,j}  k^{t   \ell_i}  \rho_j  |\braket{e_i|\rho_j}|^2  \right)$.
  Noting   that  minimizing  $\ell_t(C(\rho))$   is  equivalent   to  minimizing
  $\sum_{i,j}  k^{t \ell_i}  \rho_j |\braket{e_i|\rho_j}|^2$,  the proof  is the
  very same than that of theorem~\ref{th:optcod}, where $\vec{\ell}$ is replaced
  by $\left[k^{t  \ell_1} \ldots k^{t  \ell_d} \right]$ and where  the classical
  optimal code turns  to be $\{ c_t^\opt(i) \}$.  This last  one can be computed
  by the algorithms proposed in~\cite{Parker1980, Humblet1981, Baer2006}.
\end{proof}


Proof of theorem~\ref{th:t-bounds}.

\begin{proof}
  The proof is similar to that of theorem~\ref{th:bounds}.
  Let $C(\rho)  = U \rho  U^\dag$ with $U$  an arbitrary encoding scheme  of the
  form \eqref{eq:HayCode}. Then, noting that  $\rho U^\dag k^{t \Lambda} U =
    \rho \left(  U^\dag k^{-\Lambda}  U \right)^{-t} =  \rho_t^{1+t} \sigma^{t}
    \beta^t  \left[  \Tr\left(  \rho^{\frac{1}{1+t}} \right)  \right]^{(1+t)}$,
    where $\sigma$ and $\beta$ are defined in~\eqref{eq:sigmabeta} and $\rho_t$
  in~\eqref{eq:rhot},  we immediately obtain
  \begin{equation}
  \label{eq:t-avgent}
  \ell_t(C(\rho)) = S_{\frac{1}{1+t}}(\rho) + S_{1+t}(\rho_t \| \sigma) - \log_k
  \beta,
  \end{equation}
  where $S_{\alpha}(\rho \| \sigma)  = \frac{1}{\alpha-1} \log_k \Tr \rho^\alpha
  \sigma^{1-\alpha}$    is   the   quantum    R\'enyi   divergence    (see   \eg
  \cite{Petz1986}).  The quantum R\'enyi divergence being definite positive, and
  from \ $\log_k  \beta \leq 0$ \ due to  the quantum Kraft-McMillan inequality,
  it follows that
  \begin{equation}
  \label{eq:t-lowerbound}
  \ell_t(C(\rho)) \geq  S_{\frac{1}{1+t}}(\rho),
  \end{equation}
  for any encoding scheme $U$, in particular for the optimal one.

  In order  to prove the  upper bound, let  us now consider the  quantum Shannon
  code  $U^\sh_t =  \sum_{i=1}^d  \ketbra{c(i)}{\rho_i}$ of  the escort  density
  operator $\rho_t$, where the lengths  of the codewords $\left\{ c(i) \right\}$
  are  $\left\{ \ell_i  =  \lceil  - \log_k  {\rho_t}_i  \rceil \right\}$  being
  ${\rho_t}_i$ the  escort probabilities, eigenvalues of  $\rho_t$.  Notice that
  this code  satisfies the quantum  Kraft-McMillan inequality~\eqref{eq:QKMI} by
  construction.
  Then, from $\lceil  - \log_k {\rho_t}_i \rceil < - \log_k  {\rho_t}_i + 1$, we
  have
  \begin{equation}
  \label{eq:t-avgcbar}
  \ell_t(C^\sh_t(\rho)) = \frac{1}{t} \log_k \left( \sum_{i=1}^d \rho_i
  k^{t\lceil - \log_k {\rho_t}_i \rceil} \right) \, < \, S_{\frac{1}{1+t}}(\rho) +
  1.
  \end{equation}
  Because \ $\ell_t(C^\opt_t(\rho))  \leq \ell_t(C^\sh_t(\rho))$ \ by definition
  of  the  optimal  code,  the upper  bound  in~\eqref{eq:t-bounds}  immediately
  follows from this inequality.
\end{proof}


Proof of theorem~\ref{th:t-wrongcode}.

\begin{proof}
  Let $U^\sh_t = \sum_{i=1}^d \ketbra{c(i)}{\tau_i}$ be the quantum Shannon code
  of the escort  density operator $\tau_t$.  It is  straightforward to show that
  $\ell_t(C^\sh_t(\rho))    =    \frac{1}{t}    \log_k    \left(    \sum_{i=1}^d
    \braket{\tau_i|\rho|\tau_i} k^{t \lceil - \log_k {\tau_t}_i\rceil} \right)$.
  The bounds result  thus directly from $- \log_k {\tau_t}_i \,  \le \, \lceil -
  \log_k {\tau_t}_i \rceil  \, < \, -  \log_k {\tau_t}_i + 1$ \  together with \
  $\sum_{i=1}^d  \braket{\tau_i|\rho|\tau_i}   {\tau_t}_i^{-t}  =  \Tr\left(\rho
    \tau_t^{-t}  \right)  =   \left[  \Tr  \left(  \rho^{\frac{1}{1+t}}  \right)
  \right]^{1+t} \Tr\left(\rho_t^{1+t} \tau_t^{-t} \right)$.
\end{proof}


Proof of theorem~\ref{th:tradeoff}

\begin{proof}
Notice that for  this code, $U^{\sh}_t  =  \sum_{i=1}^d \ketbra{c(i)}{\rho_i}$,  the classical codewords  lengths can be expressed as
\begin{equation}
\label{eq:lUmin}
 \ell_i = \left\lceil - \log_k {\rho_t}_i \right\rceil = \left\lceil \frac{1}{1+t} (-\log_k \rho_i) + \frac{t}{1+t}
S_\frac{1}{1+t}(\rho) \right\rceil.
\end{equation}
Thus, we have that the average length is given by
\begin{equation}
  \label{eq:t-avg}
  \ell(C^\sh_t(\rho)) = \sum_{i=1}^d  {\rho}_i  \left\lceil - \log_k {\rho_t}_i \right\rceil = \left\lceil \frac{1}{1+t} S(\rho) + \frac{t}{1+t}
S_\frac{1}{1+t}(\rho) \right\rceil,
\end{equation}
so that the lower and upper bounds in~\eqref{eq:averageUmin} are directly obtained.
\end{proof}


\bibliography{QuantumCodingArxiv}

\begin{thebibliography}{30}%
\makeatletter
\providecommand \@ifxundefined [1]{%
 \@ifx{#1\undefined}
}%
\providecommand \@ifnum [1]{%
 \ifnum #1\expandafter \@firstoftwo
 \else \expandafter \@secondoftwo
 \fi
}%
\providecommand \@ifx [1]{%
 \ifx #1\expandafter \@firstoftwo
 \else \expandafter \@secondoftwo
 \fi
}%
\providecommand \natexlab [1]{#1}%
\providecommand \enquote  [1]{``#1''}%
\providecommand \bibnamefont  [1]{#1}%
\providecommand \bibfnamefont [1]{#1}%
\providecommand \citenamefont [1]{#1}%
\providecommand \href@noop [0]{\@secondoftwo}%
\providecommand \href [0]{\begingroup \@sanitize@url \@href}%
\providecommand \@href[1]{\@@startlink{#1}\@@href}%
\providecommand \@@href[1]{\endgroup#1\@@endlink}%
\providecommand \@sanitize@url [0]{\catcode `\\12\catcode `\$12\catcode
  `\&12\catcode `\#12\catcode `\^12\catcode `\_12\catcode `\%12\relax}%
\providecommand \@@startlink[1]{}%
\providecommand \@@endlink[0]{}%
\providecommand \url  [0]{\begingroup\@sanitize@url \@url }%
\providecommand \@url [1]{\endgroup\@href {#1}{\urlprefix }}%
\providecommand \urlprefix  [0]{URL }%
\providecommand \Eprint [0]{\href }%
\providecommand \doibase [0]{http://dx.doi.org/}%
\providecommand \selectlanguage [0]{\@gobble}%
\providecommand \bibinfo  [0]{\@secondoftwo}%
\providecommand \bibfield  [0]{\@secondoftwo}%
\providecommand \translation [1]{[#1]}%
\providecommand \BibitemOpen [0]{}%
\providecommand \bibitemStop [0]{}%
\providecommand \bibitemNoStop [0]{.\EOS\space}%
\providecommand \EOS [0]{\spacefactor3000\relax}%
\providecommand \BibitemShut  [1]{\csname bibitem#1\endcsname}%
\let\auto@bib@innerbib\@empty
\bibitem [{\citenamefont {Shannon}(1948)}]{Shannon1948}%
  \BibitemOpen
  \bibfield  {author} {\bibinfo {author} {\bibfnamefont {C.~E.}\ \bibnamefont
  {Shannon}},\ }\href {\doibase 10.1002/j.1538-7305.1948.tb00917.x} {\bibfield
  {journal} {\bibinfo  {journal} {Bell System Technical Journal}\ }\textbf
  {\bibinfo {volume} {27}},\ \bibinfo {pages} {379} (\bibinfo {year}
  {1948})}\BibitemShut {NoStop}%
\bibitem [{\citenamefont {Cover}\ and\ \citenamefont
  {Thomas}(2006)}]{CoverBook}%
  \BibitemOpen
  \bibfield  {author} {\bibinfo {author} {\bibfnamefont {T.~M.}\ \bibnamefont
  {Cover}}\ and\ \bibinfo {author} {\bibfnamefont {J.~A.}\ \bibnamefont
  {Thomas}},\ }\href@noop {} {\emph {\bibinfo {title} {Elements of information
  theory}}}\ (\bibinfo  {publisher} {John Wiley \& Sons},\ \bibinfo {year}
  {2006})\BibitemShut {NoStop}%
\bibitem [{\citenamefont {Kraft}(1949)}]{Kraft1949}%
  \BibitemOpen
  \bibfield  {author} {\bibinfo {author} {\bibfnamefont {L.~J.}\ \bibnamefont
  {Kraft}},\ }\emph {\bibinfo {title} {A device for quantizing, grouping, and
  coding amplitude-modulated pulses}},\ \href@noop {} {Ph.D. thesis},\ \bibinfo
   {school} {Massachusetts Institute of Technology} (\bibinfo {year}
  {1949})\BibitemShut {NoStop}%
\bibitem [{\citenamefont {McMillan}(1956)}]{McMillan1956}%
  \BibitemOpen
  \bibfield  {author} {\bibinfo {author} {\bibfnamefont {B.}~\bibnamefont
  {McMillan}},\ }\href {\doibase 10.1109/TIT.1956.1056818} {\bibfield
  {journal} {\bibinfo  {journal} {IRE Transactions on Information Theory}\
  }\textbf {\bibinfo {volume} {2}},\ \bibinfo {pages} {115} (\bibinfo {year}
  {1956})}\BibitemShut {NoStop}%
\bibitem [{\citenamefont {Campbell}(1965)}]{Campbell1965}%
  \BibitemOpen
  \bibfield  {author} {\bibinfo {author} {\bibfnamefont {L.}~\bibnamefont
  {Campbell}},\ }\href {\doibase
  http://dx.doi.org/10.1016/S0019-9958(65)90332-3} {\bibfield  {journal}
  {\bibinfo  {journal} {Information and Control}\ }\textbf {\bibinfo {volume}
  {8}},\ \bibinfo {pages} {423} (\bibinfo {year} {1965})}\BibitemShut {NoStop}%
\bibitem [{\citenamefont {Yamano}(2001)}]{Yamano2001}%
  \BibitemOpen
  \bibfield  {author} {\bibinfo {author} {\bibfnamefont {T.}~\bibnamefont
  {Yamano}},\ }\href {\doibase 10.1103/PhysRevE.63.046105} {\bibfield
  {journal} {\bibinfo  {journal} {Physical Review E}\ }\textbf {\bibinfo
  {volume} {63}},\ \bibinfo {pages} {046105} (\bibinfo {year}
  {2001})}\BibitemShut {NoStop}%
\bibitem [{\citenamefont {Baer}(2006)}]{Baer2006}%
  \BibitemOpen
  \bibfield  {author} {\bibinfo {author} {\bibfnamefont {M.~B.}\ \bibnamefont
  {Baer}},\ }\href {\doibase 10.1109/TIT.2006.881728} {\bibfield  {journal}
  {\bibinfo  {journal} {IEEE Transactions on Information Theory}\ }\textbf
  {\bibinfo {volume} {52}},\ \bibinfo {pages} {4380} (\bibinfo {year}
  {2006})}\BibitemShut {NoStop}%
\bibitem [{\citenamefont {Bercher}(2009)}]{Bercher2009}%
  \BibitemOpen
  \bibfield  {author} {\bibinfo {author} {\bibfnamefont {J.-F.}\ \bibnamefont
  {Bercher}},\ }\href {\doibase 10.1016/j.physleta.2009.07.015} {\bibfield
  {journal} {\bibinfo  {journal} {Physics Letters A}\ }\textbf {\bibinfo
  {volume} {373}},\ \bibinfo {pages} {3235} (\bibinfo {year}
  {2009})}\BibitemShut {NoStop}%
\bibitem [{\citenamefont {Chapeau-Blondeau}\ \emph {et~al.}(2011)\citenamefont
  {Chapeau-Blondeau}, \citenamefont {Delahaies},\ and\ \citenamefont
  {Rousseau}}]{Chapeau2011}%
  \BibitemOpen
  \bibfield  {author} {\bibinfo {author} {\bibfnamefont {F.}~\bibnamefont
  {Chapeau-Blondeau}}, \bibinfo {author} {\bibfnamefont {A.}~\bibnamefont
  {Delahaies}}, \ and\ \bibinfo {author} {\bibfnamefont {D.}~\bibnamefont
  {Rousseau}},\ }\href {\doibase 10.1049/el.2010.2792} {\bibfield  {journal}
  {\bibinfo  {journal} {Electronics Letters}\ }\textbf {\bibinfo {volume}
  {47}},\ \bibinfo {pages} {187} (\bibinfo {year} {2011})}\BibitemShut
  {NoStop}%
\bibitem [{\citenamefont {{R\'e}nyi}(1961)}]{Renyi1961}%
  \BibitemOpen
  \bibfield  {author} {\bibinfo {author} {\bibfnamefont {A.}~\bibnamefont
  {{R\'e}nyi}},\ }in\ \href@noop {} {\emph {\bibinfo {booktitle} {Proceedings
  of the fourth Berkeley symposium on mathematical statistics and
  probability}}},\ Vol.~\bibinfo {volume} {1}\ (\bibinfo {year} {1961})\ pp.\
  \bibinfo {pages} {547--561}\BibitemShut {NoStop}%
\bibitem [{\citenamefont {Campbell}(1966)}]{Campbell1966}%
  \BibitemOpen
  \bibfield  {author} {\bibinfo {author} {\bibfnamefont {L.~L.}\ \bibnamefont
  {Campbell}},\ }\href {\doibase 10.1007/BF00537132} {\bibfield  {journal}
  {\bibinfo  {journal} {Zeitschrift f{\"u}r Wahrscheinlichkeitstheorie und
  Verwandte Gebiete}\ }\textbf {\bibinfo {volume} {6}},\ \bibinfo {pages} {113}
  (\bibinfo {year} {1966})}\BibitemShut {NoStop}%
\bibitem [{\citenamefont {Schumacher}\ and\ \citenamefont
  {Westmoreland}(2001)}]{Schumacher2001}%
  \BibitemOpen
  \bibfield  {author} {\bibinfo {author} {\bibfnamefont {B.}~\bibnamefont
  {Schumacher}}\ and\ \bibinfo {author} {\bibfnamefont {M.~D.}\ \bibnamefont
  {Westmoreland}},\ }\href {\doibase 10.1103/PhysRevA.64.042304} {\bibfield
  {journal} {\bibinfo  {journal} {Physical Review A}\ }\textbf {\bibinfo
  {volume} {64}},\ \bibinfo {pages} {042304} (\bibinfo {year}
  {2001})}\BibitemShut {NoStop}%
\bibitem [{\citenamefont {Bostroem}\ and\ \citenamefont
  {Felbinger}(2002)}]{Bostroem2002}%
  \BibitemOpen
  \bibfield  {author} {\bibinfo {author} {\bibfnamefont {K.}~\bibnamefont
  {Bostroem}}\ and\ \bibinfo {author} {\bibfnamefont {T.}~\bibnamefont
  {Felbinger}},\ }\href {\doibase 10.1103/PhysRevA.65.032313} {\bibfield
  {journal} {\bibinfo  {journal} {Physical Review A}\ }\textbf {\bibinfo
  {volume} {65}},\ \bibinfo {pages} {032313} (\bibinfo {year}
  {2002})}\BibitemShut {NoStop}%
\bibitem [{\citenamefont {Koashi}\ and\ \citenamefont
  {Imoto}(2002)}]{Koashi2002}%
  \BibitemOpen
  \bibfield  {author} {\bibinfo {author} {\bibfnamefont {M.}~\bibnamefont
  {Koashi}}\ and\ \bibinfo {author} {\bibfnamefont {N.}~\bibnamefont {Imoto}},\
  }\href {\doibase 10.1103/PhysRevLett.89.097904} {\bibfield  {journal}
  {\bibinfo  {journal} {Physical ReviewLetters}\ }\textbf {\bibinfo {volume}
  {89}},\ \bibinfo {pages} {097904} (\bibinfo {year} {2002})}\BibitemShut
  {NoStop}%
\bibitem [{\citenamefont {Ahlswede}\ and\ \citenamefont
  {Cai}(2003)}]{Ahlswede2003}%
  \BibitemOpen
  \bibfield  {author} {\bibinfo {author} {\bibfnamefont {R.}~\bibnamefont
  {Ahlswede}}\ and\ \bibinfo {author} {\bibfnamefont {N.}~\bibnamefont {Cai}},\
  }in\ \href@noop {} {\emph {\bibinfo {booktitle} {Quantum Information
  Processing}}},\ \bibinfo {editor} {edited by\ \bibinfo {editor}
  {\bibfnamefont {G.}~\bibnamefont {Leuchs}}\ and\ \bibinfo {editor}
  {\bibfnamefont {T.}~\bibnamefont {Beth}}}\ (\bibinfo  {publisher}
  {Wiley-VCH},\ \bibinfo {year} {2003})\ Chap.~\bibinfo {chapter} {6}, pp.\
  \bibinfo {pages} {66--78}\BibitemShut {NoStop}%
\bibitem [{\citenamefont {Ahlswede}\ and\ \citenamefont
  {Cai}(2004)}]{Ahlswede2004}%
  \BibitemOpen
  \bibfield  {author} {\bibinfo {author} {\bibfnamefont {R.}~\bibnamefont
  {Ahlswede}}\ and\ \bibinfo {author} {\bibfnamefont {N.}~\bibnamefont {Cai}},\
  }\href {\doibase 10.1109/TIT.2004.828071} {\bibfield  {journal} {\bibinfo
  {journal} {IEEE Transactions on Information Theory}\ }\textbf {\bibinfo
  {volume} {50}},\ \bibinfo {pages} {1208} (\bibinfo {year}
  {2004})}\BibitemShut {NoStop}%
\bibitem [{\citenamefont {M{\"u}ller}\ and\ \citenamefont
  {Rogers}(2008)}]{Muller2008}%
  \BibitemOpen
  \bibfield  {author} {\bibinfo {author} {\bibfnamefont {M.}~\bibnamefont
  {M{\"u}ller}}\ and\ \bibinfo {author} {\bibfnamefont {C.}~\bibnamefont
  {Rogers}},\ }\href {https://arxiv.org/abs/0804.0022} {\bibfield  {journal}
  {\bibinfo  {journal} {arXiv preprint arXiv:0804.0022}\ } (\bibinfo {year}
  {2008})}\BibitemShut {NoStop}%
\bibitem [{\citenamefont {M{\"u}ller}\ \emph {et~al.}(2009)\citenamefont
  {M{\"u}ller}, \citenamefont {Rogers},\ and\ \citenamefont
  {Nagarajan}}]{Muller2009}%
  \BibitemOpen
  \bibfield  {author} {\bibinfo {author} {\bibfnamefont {M.}~\bibnamefont
  {M{\"u}ller}}, \bibinfo {author} {\bibfnamefont {C.}~\bibnamefont {Rogers}},
  \ and\ \bibinfo {author} {\bibfnamefont {R.}~\bibnamefont {Nagarajan}},\
  }\href {\doibase 10.1103/PhysRevA.79.012302} {\bibfield  {journal} {\bibinfo
  {journal} {Physical Review A}\ }\textbf {\bibinfo {volume} {79}},\ \bibinfo
  {pages} {012302} (\bibinfo {year} {2009})}\BibitemShut {NoStop}%
\bibitem [{\citenamefont {Hayashi}(2010)}]{Hayashi2010}%
  \BibitemOpen
  \bibfield  {author} {\bibinfo {author} {\bibfnamefont {M.}~\bibnamefont
  {Hayashi}},\ }\href {\doibase 10.1007/s00220-009-0909-y} {\bibfield
  {journal} {\bibinfo  {journal} {Communications in Mathematical Physics}\
  }\textbf {\bibinfo {volume} {293}},\ \bibinfo {pages} {171} (\bibinfo {year}
  {2010})}\BibitemShut {NoStop}%
\bibitem [{\citenamefont {Hayashi}(2017{\natexlab{a}})}]{HayashiBook}%
  \BibitemOpen
  \bibfield  {author} {\bibinfo {author} {\bibfnamefont {M.}~\bibnamefont
  {Hayashi}},\ }\href {\doibase 10.1007/978-3-319-45241-8} {\emph {\bibinfo
  {title} {A Group Theoretic Approach to Quantum Information}}}\ (\bibinfo
  {publisher} {Springer},\ \bibinfo {address} {Berlin},\ \bibinfo {year}
  {2017})\BibitemShut {NoStop}%
\bibitem [{\citenamefont {Huffman}(1952)}]{Huffman1952}%
  \BibitemOpen
  \bibfield  {author} {\bibinfo {author} {\bibfnamefont {D.~A.}\ \bibnamefont
  {Huffman}},\ }\href {\doibase 10.1109/JRPROC.1952.273898} {\bibfield
  {journal} {\bibinfo  {journal} {Proceedings of the IRE}\ }\textbf {\bibinfo
  {volume} {40}},\ \bibinfo {pages} {1098} (\bibinfo {year}
  {1952})}\BibitemShut {NoStop}%
\bibitem [{\citenamefont {Schumacher}\ and\ \citenamefont
  {Westmoreland}(2002)}]{SchumacherBook}%
  \BibitemOpen
  \bibfield  {author} {\bibinfo {author} {\bibfnamefont {B.}~\bibnamefont
  {Schumacher}}\ and\ \bibinfo {author} {\bibfnamefont {M.~D.}\ \bibnamefont
  {Westmoreland}},\ }\href@noop {} {\bibfield  {journal} {\bibinfo  {journal}
  {Contemporary Mathematics}\ }\textbf {\bibinfo {volume} {305}},\ \bibinfo
  {pages} {265} (\bibinfo {year} {2002})}\BibitemShut {NoStop}%
\bibitem [{\citenamefont {Kaltchenko}(2008)}]{Kaltchenko2008}%
  \BibitemOpen
  \bibfield  {author} {\bibinfo {author} {\bibfnamefont {A.}~\bibnamefont
  {Kaltchenko}},\ }\href {\doibase 10.1103/PhysRevA.78.022311} {\bibfield
  {journal} {\bibinfo  {journal} {Physical Review A}\ }\textbf {\bibinfo
  {volume} {78}},\ \bibinfo {pages} {022311} (\bibinfo {year}
  {2008})}\BibitemShut {NoStop}%
\bibitem [{\citenamefont {Parker}(1980)}]{Parker1980}%
  \BibitemOpen
  \bibfield  {author} {\bibinfo {author} {\bibfnamefont {D.~S.}\ \bibnamefont
  {Parker}},\ }\href {\doibase 10.1137/0209035} {\bibfield  {journal} {\bibinfo
   {journal} {SIAM Journal on Computing}\ }\textbf {\bibinfo {volume} {9}},\
  \bibinfo {pages} {470} (\bibinfo {year} {1980})}\BibitemShut {NoStop}%
\bibitem [{\citenamefont {Humblet}(1981)}]{Humblet1981}%
  \BibitemOpen
  \bibfield  {author} {\bibinfo {author} {\bibfnamefont {P.~A.}\ \bibnamefont
  {Humblet}},\ }\href {\doibase 10.1109/TIT.1981.1056322} {\bibfield  {journal}
  {\bibinfo  {journal} {IEEE Transactions on Information Theory}\ }\textbf
  {\bibinfo {volume} {27}},\ \bibinfo {pages} {230} (\bibinfo {year}
  {1981})}\BibitemShut {NoStop}%
\bibitem [{\citenamefont {Hayashi}(2002)}]{Hayashi2002}%
  \BibitemOpen
  \bibfield  {author} {\bibinfo {author} {\bibfnamefont {M.}~\bibnamefont
  {Hayashi}},\ }\href {\doibase 10.1103/PhysRevA.66.032321} {\bibfield
  {journal} {\bibinfo  {journal} {Phys. Rev. A}\ }\textbf {\bibinfo {volume}
  {66}},\ \bibinfo {pages} {032321} (\bibinfo {year} {2002})}\BibitemShut
  {NoStop}%
\bibitem [{\citenamefont {Hayashi}(2017{\natexlab{b}})}]{HayashiBook2}%
  \BibitemOpen
  \bibfield  {author} {\bibinfo {author} {\bibfnamefont {M.}~\bibnamefont
  {Hayashi}},\ }\href {\doibase 10.1007/978-3-662-49725-8} {\emph {\bibinfo
  {title} {Quantum Information Theory}}}\ (\bibinfo  {publisher} {Springer},\
  \bibinfo {address} {Berlin},\ \bibinfo {year} {2017})\BibitemShut {NoStop}%
\bibitem [{\citenamefont {Birkhoff}(1946)}]{Birkhoff1946}%
  \BibitemOpen
  \bibfield  {author} {\bibinfo {author} {\bibfnamefont {G.}~\bibnamefont
  {Birkhoff}},\ }\href@noop {} {\bibfield  {journal} {\bibinfo  {journal}
  {Universidad Nacional de {T}ucum{\'a}n. Revista. Serie A, mat{\'e}matica y
  f{\'i}sica te{\'o}rica}\ }\textbf {\bibinfo {volume} {5}},\ \bibinfo {pages}
  {147} (\bibinfo {year} {1946})}\BibitemShut {NoStop}%
\bibitem [{\citenamefont {Bhatia}(1997)}]{Bhatia1997}%
  \BibitemOpen
  \bibfield  {author} {\bibinfo {author} {\bibfnamefont {R.}~\bibnamefont
  {Bhatia}},\ }\href@noop {} {\emph {\bibinfo {title} {Matrix Analysis}}}\
  (\bibinfo  {publisher} {Springer Verlag},\ \bibinfo {address} {New-York},\
  \bibinfo {year} {1997})\BibitemShut {NoStop}%
\bibitem [{\citenamefont {Petz}(1986)}]{Petz1986}%
  \BibitemOpen
  \bibfield  {author} {\bibinfo {author} {\bibfnamefont {D.}~\bibnamefont
  {Petz}},\ }\href {\doibase http://dx.doi.org/10.1016/0034-4877(86)90067-4}
  {\bibfield  {journal} {\bibinfo  {journal} {Reports on Mathematical Physics}\
  }\textbf {\bibinfo {volume} {23}},\ \bibinfo {pages} {57} (\bibinfo {year}
  {1986})}\BibitemShut {NoStop}%
\end{thebibliography}%


\section*{Acknowledgements}

The authors acknowledge CONICET and UNLP (Argentina) and CNRS (France) for partial support.










\end{document}